\documentclass{aa}
\usepackage{graphicx}
\begin{document}

\title{On the oxygen abundance in our Galaxy}

\author{L.S.~Pilyugin \inst{1}, 
        Federico Ferrini \inst{2},
        R.V.~Shkvarun \inst{1}
        }

\offprints{L.S. Pilyugin }

\institute{Main Astronomical Observatory
           of National Academy of Sciences of Ukraine,
           27 Zabolotnogo str., 03680 Kiev, Ukraine, 
           (pilyugin@mao.kiev.ua, shkvarun@mao.kiev.ua)
           \and
           Department of Physics, Section of Astronomy,
           University of Pisa, piazza Torricelli 2,
           56100 Pisa, Italy, \\   
           (federico.ferrini@df.unipi.it)                 }

\date{Received 12 November 2002 / accepted 9 January 2003}

\abstract{
The compilation of published spectra of Galactic H\,{\sc ii} regions with 
available diagnostic [OIII]$\lambda$4363 line information has been carried out. 
Our list contains 71 individual measurements of 13 H\,{\sc ii} regions in the 
range of galactocentric distances from 6.6 to 14.8 kpc. The oxygen abundances in 
all the H\,{\sc ii} regions were recomputed in the same way, using the classic 
T$_e$ - method. The oxygen abundance at the solar galactocentric distance 
traced by those H\,{\sc ii} regions is in agreement with the oxygen abundance 
in the interstellar medium in the solar vicinity derived with high precision 
from the interstellar absorption lines towards stars. The derived radial oxygen 
abundance distribution was compared with that for H\,{\sc ii} regions from the 
Shaver et al. (1983) sample which is the basis of many models for the chemical 
evolution of our Galaxy. It was found that the original Shaver et al.'s oxygen 
abundances are overestimated by 0.2-0.3 dex. Oxygen abundances in H\,{\sc ii} 
regions from the Shaver et al. sample have been redetermined with the recently 
suggested P -- method. The radial distribution of oxygen abundances from the 
Shaver et al. sample redetermined with the P -- method is in agreement with our 
radial distribution of (O/H)$_{T_{e}}$ abundances.
}

\titlerunning{On the oxygen abundance in our Galaxy}

\authorrunning{L.S.~Pilyugin et al.}

\maketitle

\section{Introduction}

By now, spectra have been obtained for hundreds of H\,{\sc ii} regions in spiral 
galaxies. Accurate oxygen abundances can be derived from measurement of 
temperature-sensitive line ratios, such as [OIII]4959,5007/[OIII]4363. This 
method will be referred to as the T$_{e}$ - method. Unfortunately, in 
oxygen-rich H\,{\sc ii} regions the temperature-sensitive lines such as [OIII]4363 are 
too weak to be detected. For such H\,{\sc ii} regions, empirical abundance indicators 
based on more readily observable lines were suggested (Pagel et al. 1979; Alloin 
et al. 1979). The empirical oxygen abundance indicator R$_{23}$ = 
([OII]3727,3729 + [OIII]4959,5007)/H$_{\beta}$, suggested by Pagel et al. (1979), 
has found widespread acceptance and use for the oxygen abundance determination 
in H\,{\sc ii} regions where the temperature-sensitive lines are undetectable. 
This method will be referred to as the R$_{23}$ - method. Using the R$_{23}$ - 
method, the characteristic oxygen abundances (the oxygen abundance at a 
predetermined galactocentric distance) and radial oxygen abundance gradients 
were obtained for a large sample of spiral galaxies (Vila-Costas \& Edmunds 1992; 
Zaritsky et al. 1994; van Zee et al. 1998, among others).

It has been found (Pilyugin 2001b) that the (O/H)$_{T_{e}}$ data is sufficient 
in quantity and quality for an accurate determination of the value of the 
oxygen abundance gradient within the disk of the spiral galaxy M101. 
It has been found that the parameters of the (O/H)$_{R_{23}}$ abundance 
distribution in the disk of M101 differ significantly from those
of the (O/H)$_{T_{e}}$ abundance distribution. 
The rather low value of the central oxygen abundance 
12 + log(O/H)$_{T_{e}}$(R=0) = 8.81 in the disk of M101 was found. This is in 
agreement with the result of Kinkel \& Rosa (1994). They carried out special 
spectral observastions of one single high-metallicity H\,{\sc ii} region in the disk 
of the spiral galaxy M101 (NGC5457) and were able to derive the electron 
temperature, allowing the determination of the oxygen abundance with the 
classic T$_{e}$ - method. Kinkel \& Rosa (1994) found that the oxygen abundance 
derived with the T$_{e}$ - method is lower by around 0.2 dex with respect to the 
oxygen abundance obtained on the basis of the R$_{23}$ calibration after 
Edmunds \& Pagel (1984). Castellanos et al. (2002) have also found that the 
R$_{23}$ - method results in the overestimated oxygen abundances in the case of 
low-excitation H\,{\sc ii} regions. 

Pilyugin et al. (2002) have re-determined the oxygen abundances in H\,{\sc ii} 
regions in a number of field and Virgo spiral galaxies with the P -- method. 
It has been found for the considered sample of galaxies that the central 
values 12 + log(O/H)$_{P}$(R=0) lie in the range 8.71 $\div$ 8.99 while the 
central values 12 + log(O/H)$_{R_{23}}$(R=0) lie in the range 9.24 $\div$ 9.53. 
Thus, for all the galaxies, in which the oxygen abundances in H\,{\sc ii} 
regions were derived with the $T_{e}$ -- or with the P -- method, the value of 
the central oxygen abundance does not exceed 12 + log(O/H)(R=0) $\sim$ 9.0. 

The radial distribution of oxygen and other element abundances in our Galaxy 
derived by Shaver et al. (1983) from consideration of H\,{\sc ii} regions has 
found wide acceptance. The Shaver et al.'s data are the basis of the models for 
the chemical evolution of the Milky Way Galaxy constructed by different 
investigators (Tosi 1988; Giovagnoli \& Tosi 1995; Thon \& Meusinger 1998, 
among others). Shaver et al. (1983) have found the value of the central oxygen 
abundance in our Galaxy as large as 12 + log(O/H)(R=0) = 9.38$\pm$0.04. 
Radio recombination lines have been used by Shaver et al. to determine 
electron temperatures in H\,{\sc ii} regions and these temperatures have been 
applied to optical spectra of the same H\,{\sc ii} regions to determine the 
abundances of oxygen and other elements. The Galactic H\,{\sc ii} regions with  
measured temperature-sensitive line ratios will be considered in the present 
study, and, as consequence, the radial distribution of the (O/H)$_{T_{e}}$ 
abundances within the disk of the Milky Way Galaxy will be established. The 
comparison of the obtained radial distribution of the (O/H)$_{T_{e}}$ abundances 
with that from Shaver et al. (1983) allows us to verify the credibility of the 
oxygen abundances derived by Shaver et al. (1983).

\section{H\,{\sc ii} regions with available [OIII]$\lambda$4363 line intensity}

Spectroscopic observations of Galactic H\,{\sc ii} regions have been carried 
out by many investigators (Baldwin et al. 1991, 2000; Caplan et al. 2000; 
Peimbert et al. 1969, 1977, 1978, 1992, 1993; Esteban 1999; Esteban et al. 1998, 
1999;  Osterbrock et al. 1992; Shaver et al. 1983; V\'{\i}lchez \& Esteban 1996, 
among others). The Galactic H\,{\sc ii} regions with  measured 
temperature-sensitive line ratios are listed in Table \ref{table:tesample}. 

Spectroscopic data with detections of diagnostic emission lines in the H\,{\sc ii} 
region makes it possible to determine an accurate oxygen abundance (O/H)$_{T_{e}}$. 
However, the oxygen abundances in the same H\,{\sc ii} region with measured line ratios 
[OIII]$\lambda \lambda 4959, 5007 / \lambda 4363$ derived in different works can 
differ due to different atomic data adopted and different interpretation of 
the temperature structure (e.g. single characteristic $T_{e}$, two-zone model for 
$T_{e}$, model with small-scale temperature fluctuations). 
Therefore the compilation of H\,{\sc ii} regions with original oxygen 
abundance determinations through the T$_{e}$ -- method from different works 
carried out over more than thirty years is not a set of homogeneous 
determinations. Accordingly, the available published spectra of H\,{\sc ii} regions with 
measured line ratios [OIII]$\lambda \lambda 4959, 5007 / \lambda 4363$ 
have been reanalysed to produce a homogeneous set. Two-zone models of H\,{\sc ii} 
regions with the algorithm for oxygen abundance determination from Pagel et al. 
(1992) and the T$_{e}$([OII]) -- T$_{e}$([OIII]) relation from Garnett (1992) were 
adopted here, 
\begin{equation}
\frac{O}{H} = \frac{O^+}{H^+} + \frac{O^{++}}{H^+}                ,
\label{equation:otot}
\end{equation}
\begin{eqnarray}
12+ \log (O^{++}/H^+) = \log \frac{I_{[OIII] \lambda 4959 + \lambda 5007}}
{I_{H_{\beta}}} + 6.174 +  \nonumber  \\ 
\frac{1.251}{t_3}  - 0.55 \log t_3 ,
\label{equation:oplus2}
\end{eqnarray}
\begin{eqnarray}
12+ \log (O^{+}/H^+) = \log \frac{I_{[OII] \lambda 3726 + \lambda 3729}}
{I_{H_{\beta}}} + 5.890 +  \nonumber  \\ 
\frac{1.676}{t_2}  - 0.40 \log t_2 + \log (1+1.35x)  ,
\label{equation:oplus}
\end{eqnarray}
\begin{eqnarray}
t_3 = t([OIII]) = 1.432 [ \log \frac{I_{[OIII] \lambda 4959 + \lambda 5007}}
{I_{[OIII] \lambda 4363}} - 0.85 + \nonumber  \\ 
0.03 \log t_3  +  \log (1 + 0.0433 x t_3^{0.06})]^{-1} ,
\label{equation:t3}
\end{eqnarray}
\begin{equation}
x= 10^{-4} n_e t_2^{-1/2}, 
\end{equation}
where $n_e$ is the electron density in cm$^{-3}$ (the value of $n_e$ was adopted 
to be equal to 100 cm$^{-3}$ for every H\,{\sc ii} region), $t_3$ = $t_{[OIII]}$ 
is the electron temperature within the [OIII] zone in units of 10$^4$K, $t_2$ = 
$t_{[OII]}$ is the electron temperature within the [OII] zone in units of 10$^4$K. 
The $t_2$ value is determined from the equation (Garnett 1992)
\begin{equation}
t_2 =  0.7 \, t_3 + 0.3 .
\label{equation:t2t3}
\end{equation}

\begin{table}
\caption[]{\label{table:tesample}
Oxygen abundance (mean value $\langle (O/H)_{Te} \rangle$) in Galactic 
H\,{\sc ii} regions with available [OIII]$\lambda$4363 line intensity.
{\it List of references}:
 1 -- Esteban et al. (1999); 
 2 -- Peimbert \& Costero (1969);
 3 -- Peimbert et al. (1993);
 4 -- Caplan et al. (2000);
 5 -- Deharveng et al. (2000);
 6 -- Esteban (1999);
 7 -- Peimbert et al. (1992);
 8 -- Peimbert et al. (1978);
 9 -- Baldwin et al. (1991);
10 -- Baldwin et al. (2000); 
11 -- Esteban et al. (1998);
12 -- Osterbrock et al. (1992);
13 -- Peimbert et al. (1977);
14 -- Shaver et al. (1983); 
15 -- V\'{\i}lchez \& Esteban (1996)
}
\begin{center}
\begin{tabular}{lccll} \hline \hline
               & galacto-    &  number    &            &               \\
H\,{\sc ii} region     & centric     &  of        & ref        &(O/H)$_{Te}^*$ \\  
               & distance    &  measu-    &            &               \\
               &  kpc        &  rements   &            &               \\  \hline
M8             &  6.6        &   3        & 1,2,3      & 8.48$\pm$0.06 \\          
M16            &  6.6        &   1        & 4,5        & 8.58          \\          
M17            &  7.1        &   18       & 2,6,7      & 8.60$\pm$0.05 \\          
Carina         &  8.1        &   5        & 8          & 8.36$\pm$0.03 \\          
Sh117          &  8.5        &   2        & 4,5        & 8.58$\pm$0.08 \\          
Orion          &  8.8        &   24       & 2,4,5,9-13 & 8.49$\pm$0.06 \\          
Sh184          &  10.1       &   4        & 4,5        & 8.48$\pm$0.02 \\          
Sh311          &  10.9       &   3        & 8,14       & 8.26$\pm$0.20 \\          
Sh206          &  11.1       &   4        & 4,5        & 8.47$\pm$0.03 \\          
Sh128          &  12.7       &   1        & 15         & 8.04$\pm$0.20$^1$ \\          
Sh298          &  13.0       &   3        & 8,14       & 8.26$\pm$0.01 \\          
Sh127          &  13.9       &   1        & 15         & 8.06$\pm$0.20$^1$ \\          
Sh212          &  14.8       &   2        & 4,5        & 8.32$\pm$0.04 \\  \hline
\end{tabular}
\end{center}

$^*$  in units of 12+log(O/H)

$^1$  original data
 
\end{table}

\begin{figure}
\centering
\resizebox{1.00\hsize}{!}{\includegraphics[angle=0]{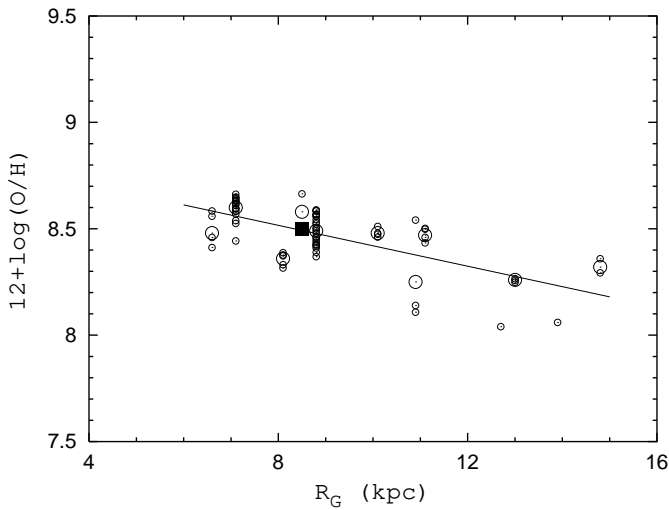}}
\caption{
Oxygen abundance in Galactic H\,{\sc ii} regions with an available [OIII]$\lambda$4363 
line intensity as a function of galactocentric distance. Small open circles
are oxygen abundances $(O/H)_{Te}$ for individual measurements, the solid 
line is the best fit to those data. Large open circles are mean values of
oxygen abundances $\langle (O/H)_{Te} \rangle$ in H\,{\sc ii} regions. 
The large filled square is the oxygen abundance in the interstellar gas in 
the solar vicinity derived from the high-resolution observations of the weak 
interstellar OI$\lambda$1356 absorption lines towards stars. 
}
\label{figure:ohte}
\end{figure}

The electron temperatures T$_{e}$([OIII]) in H\,{\sc ii} regions (with two 
exceptions, H\,{\sc ii} regions S127 and S128) have been determined from the 
measurements of [OIII]$\lambda \lambda 4959, 5007 / \lambda 4363$ line ratios, 
and the electron temperatures T$_{e}$([OII]) in H\,{\sc ii} regions have been 
derived from the T$_{e}$([OII]) -- T$_{e}$([OIII]) relation of Garnett (1992). 
In the case of H\,{\sc ii} regions S127 and S128 the electron temperatures 
T$_{e}$([OIII]) cannot be directly determined from observational data since the 
measurement of the [OIII]$\lambda \lambda 4959, 5007 / \lambda 4363$ line ratio 
is not available. Instead, the temperature-sensitive lines [OII]$\lambda \lambda 7320, 7330$ 
were detected in spectrophotometry of H\,{\sc ii} regions S127 and S128
(V\'{\i}lchez \& Esteban 1996). The original oxygen abundances in these H\,{\sc ii} 
regions are given in Table \ref{table:tesample}.

In Fig.\ref{figure:ohte} we show oxygen abundances (O/H)$_{T_{e}}$ for the 
individual measurements as a function of galactocentric distance with small 
open circles. The galactocentric distances are from Deharveng et al. (2000). 
The best fit to the individual (O/H)$_{T_{e}}$ data is
\begin{equation}
12 + \log (O/H)_{Te} = 8.90 - 0.048 \, R_G .
\label{equation:gradohte}
\end{equation}
The scatter is equal to 0.10 dex. Eq.(\ref{equation:gradohte}) results in the 
oxygen abundance at a solar galactocentric distance as large as 12+log(O/H) 
$\simeq$ 8.50. The (O/H)$_{Te}$ -- R$_G$ relationship is shown in Fig. \ref{figure:ohte} 
by a solid line. The large open circles in Fig.\ref{figure:ohte} are average 
oxygen abundances (O/H)$_{T_{e}}$ for H\,{\sc ii} regions (the data from 
Table \ref{table:tesample}). 

Thus, the obtained value of the central oxygen abundance in the disk of the 
Milky Way Galaxy 12 + log(O/H)$_{Te}$(R=0) = 8.90 lies in the same range as 
the values of the central oxygen abundance in the disks of other spiral 
galaxies, in which the oxygen abundances in H\,{\sc ii} regions were derived 
with the $T_{e}$ -- or with the P -- method. 

\section{The oxygen abundance in the solar vicinity}

The above obtained radial distributions of $(O/H)_{Te}$ (Eq.\ref{equation:gradohte}) 
results in the oxygen abundance at a solar galactocentric distance as large 
as 12+log(O/H) $\simeq$ 8.50. Let us compare this value with recent determinations 
by other investigators. 

Caplan et al. (2000) and Deharveng et al. (2000) have analysed Galactic 
H\,{\sc ii} regions and have obtained the slope --0.0395 dex/kpc with a central 
oxygen abundance 12+log(O/H) = 8.82 and 12+log(O/H) = 8.48 at the solar 
galactocentric distance. Rodr\'{\i}guez (1999) considered seven bright Galactic 
H\,{\sc ii} regions with galactocentric distances in the range 6 -- 10 kpc and 
has found that all the H\,{\sc ii} regions studied are characterized by similar 
abundances, 12+log(O/H) $\sim$ 8.45$\pm$0.1. Thus, our value of the oxygen 
abundance at the solar galactocentric distance is in agreement with that 
derived recently in other studies. 

High-resolution observations of the weak interstellar OI$\lambda$1356 absorption 
towards stars allow one to determine interstellar gas-phase oxygen 
abundance in the solar vicinity with very high precesion. These observations 
yield a mean interstellar gas-phase oxygen abundance of 319 O atoms per 
10$^6$ H atoms (or 12+log(O/H) = 8.50) (Meyer et al. 1998; Sofia \& Meyer 2001).  
The interstellar gas-phase oxygen abundance in the solar vicinity is shown 
in Fig.\ref{figure:ohte} by the large filled square. There are no statistically 
significant variations in the measured oxygen abundances from line of sight to 
line of sight; the rms scatter value for these oxygen abundances is low, $\pm$0.05dex. 
Out to 1.5 kpc, the gas-phase oxygen abundances are stable in diffuse clouds 
with different physical conditions as measured by the fraction of H in the 
form of H$_2$. Thus, our value of the oxygen abundance at the solar galactocentric 
distance derived from consideration of the H\,{\sc ii} regions is in agreement 
with that derived with high precision from the interstellar absorption towards 
the stars, Fig.\ref{figure:ohte}. 

The oxygen abundances in H\,{\sc ii} regions were determined within the 
framework of two-zone model for $T_{e}$ within the H\,{\sc ii} region.
The agreement between the value of the oxygen abundance at the solar 
galactocentric distance derived from consideration of the H\,{\sc ii} regions and that 
derived from the interstellar absorption towards stars can be considered as 
strong evidence in favor that the two-zone model for $T_{e}$ is a realistic 
interpretation of the temperature structure within H\,{\sc ii} regions. 
It should be noted however that this argument is not indisputable due to 
the dust-phase oxygen abundance. Indeed, one can expect that total (gas- plus
dust-phase) oxygen abundances in the H\,{\sc ii} regions are equal to the total 
oxygen abundance in the interstellar medium.  
The gas-phase oxygen abundances in the H\,{\sc ii} regions are expected to be 
equal to the gas-phase oxygen abundance in the interstellar medium if only the 
same fraction of oxygen is incorporated into dust grains both in H\,{\sc ii} 
regions and in the interstellar medium. Meyer et al. (1998) have obtained the 
limit to the dust-phase oxygen abundance in the interstellar medium in the 
vicinity of the Sun. Assuming various mixtures of oxygen-bearing 
grain compounds, they found that it is difficult to increase the oxygen dust 
fraction beyond 120 O atoms per 10$^6$ H atmos, simply because the requisite 
metals are far less abundant than oxygen (if Si/O, Mg/O and Fe/O abundance 
ratios in the present-day interstellar medium are the same as they are 
in the Sun).  Then the upper value of total (gas- plus dust-phase) oxygen 
abundance in the interstellar medium in the vicinity of the Sun is of 439 O 
atoms per million H atoms (or 12+log(O/H) = 8.64). Then, in the case of the 
dust-free H\,{\sc ii} regions the gas-phase oxygen abundances in the H\,{\sc ii} 
regions in the vicinity of the Sun would be equal to the total oxygen abundance 
and would be as large as 12+log(O/H) = 8.64. In this case the gas-phase oxygen 
abundances in H\,{\sc ii} regions, determined within the framework of two-zone 
model for $T_{e}$, would be underestimated by 0.14 dex. Taking into account that 
H\,{\sc ii} regions are not dust-free (according to Esteban et al. (1998), the 
fraction of the dust-phase oxygen abundance in the Orion nebula is about 0.1 dex)
one can expect that the gas-phase oxygen abundances in the H\,{\sc ii} regions 
are close (or equal) to the gas-phase oxygen abundance in the interstellar 
medium, and, consequently, one can conclude that two-zone model for $T_{e}$ 
distribution within the H\,{\sc ii} region provides a realistic gas-phase 
oxygen abundance in H\,{\sc ii} regions.

It has been known for a long time that permitted lines in H\,{\sc ii} regions 
indicate higher oxygen abundances than forbidden lines. Esteban et al. (1998) 
found oxygen abundances for two positions in the Orion nebula from permitted and 
forbidden lines. The oxygen abundances derived from forbidden lines are 
coincident for both positions in the Orion nebula (12 + logO/H = 8.47) and agree 
well with the interstellar oxygen abundance in the solar vicinity derived from 
interstellar absorption lines towards the stars (12 + logO/H = 8.50). They found 
from permitted lines the oxygen abundance 12 + logO/H = 8.61 for position 1 and 
12 + logO/H = 8.68 for position 2, although the abundances obtained from the 
different multiplets observed show significant dispersion. If permitted lines 
indicate the true oxygen abundance in the Orion nebula, then the uncertainty 
around 0.1 dex in the oxygen abundances derived from forbidden lines cannot be 
excluded. However, the origin of the discrepancy between abundances derived from 
permitted and forbidden lines is not indisputable (see discussion in Stasinska (2002)). 

Hence the value of total (gas- plus dust-phase) oxygen abundance in the 
interstellar medium in the vicinity of the Sun is 12+log(O/H) = 8.6$\pm$0.1. 
Young stars provide an excellent method for probing the metallicity and recent 
evolutionary history of the Milky Way Galaxy. Early B-type dwarfs have 
evolutionary ages typically less than 30 Myr and possess photospheres that are 
not normally contaminated by core-processed material. Hence, these stars should 
provide information on the present-day chemical composition of the interstellar 
medium within the Milky Way Galaxy. There have been several attempts to measure 
the oxygen abundance using young, B-type stars as tracers of metallicity in the 
Galactic disk (Fitzsimmons, Dufton \& Rolleston 1992; Kaufer et al. 1994; 
Smartt \& Rolleston 1977; Gummersbach et al. 1998; Daflon, Cunha \& Becker 1999; 
Rolleston et al. 2000; Daflon et al. 2001, among others). Unfortunately, 
results from stars are controversial. Recently, Rolleston et al. (2000) and 
Daflon et al. (2001) have derived oxygen abundances in early type stars using 
both LTE and non-LTE calculations (see the discussion of earlier results in 
Rolleston et al. (2000)). Rolleston et al. (2000) have 
found from early type stars a 12+logO/H value of around 8.80 for the solar 
vicinity, while Daflon et al. (2001) have found from a non-LTE analysis a 
12+logO/H value around 8.50 for the solar vicinity region. It should be taken 
into account however that Rolleston et al. (2000) have noted that their data 
should be reliable for estimating the magnitude of the metal abundance gradients 
in the present-day Galactic disk; although, the absolute abundance values at any 
particular galactocentric distance may be in error.

Andrievsky et al. (2002) have derived the radial distribution of the oxygen 
(and other chemical elements) abundance within the Galactic disk based on the 
spectroscopic analysis of a sample of classic Cepheids. They have 
found from Cepheids an oxygen abundance gradient in the Galactic disk that 
results in a [O/H] value of around --0.07 dex for the solar vicinity region. 
Andrievsky et al. (2002) have derived a differential abundances in the sence 
that the oscillator strengths used in their study were obtained through an 
inverted solar analysis (with adopted solar abundances). As consequence, their 
absolute oxygen abundance values depend on the adopted solar oxygen abundance.
The recent determinations of the solar oxygen abundance 
result in the value of 12+log(O/H)$_{\odot}$ $\sim$ 8.70 (Holweger 2001; Prieto 
et al. 2001), i.e. the present-day "recommended" solar abundance is less
(by more than 0.2 dex) than the value recommended several years ago (Anders \& 
Grevesse 1989;  Grevesse, Noels \& Sauval 1996; Grevesse \& Sauval 1998). 
With this recent solar oxygen abundance, the data of Andrievsky et al. (2002) 
result in a 12+logO/H value of around 8.60 for the solar vicinity. 

Thus, the value of the oxygen abundance at the solar galactocentric distance 
(12+logO/H$\sim$8.6) derived by Andrievsky et al. (2002) from the study of 
Cepheids and the value of the oxygen abundance (12+logO/H$\sim$8.5) derived by 
Daflon et al. (1999, 2001) from the study of OB stars are in agreement with 
the oxygen abundance in the solar vicinity region derived from the interstellar 
absorption towards stars and from the study of H\,{\sc ii} regions. 
At the same time, there is an appreciable disagreement between values of the 
radial oxygen abundance gradient derived from the study of Cepheids 
(--0.022dex/kpc, Andrievsky et al. 2002), from the study of early B-type 
main-sequence stars (--0.067dex/kpc, Rolleston et al. 2000), and from the 
study of H\,{\sc ii} regions (--0.048dex/kpc, this study). 
This disagreement can be explained by two reasons. First, there is a 
relatively large scatter of oxygen abundance values at any particular 
galactocentric distance for objects of any kind (Cepheids, OB stars, and 
H\,{\sc ii} regions). Second. A majority of data-points, used in the determination 
of the value of the radial oxygen abundance gradient, are distributed over a 
restricted (6 $<$ R$_G$ $<$ 13 kpc) range of galactocentric distances: all the 
Cepheids from the sample of Andrievsky et al. have galactocentric distances in the 
range 6 -- 11 kpc, only three objects from the sample of Rolleston et al. exist 
for galactocentric distances greater than R$_G$ $\sim$ 13 kpc, only three 
individual measurements of two H\,{\sc ii} regions with galactocentric distances 
greater than R$_G$ $\sim$ 13 kpc were used in this study. This disagreement 
prevents us from reaching a definitive conclusion about the value of the 
radial oxygen abundance gradient within the Galactic disk; though we stress 
that more and better data are needed to derive the exact value of gradient.

It should be noted that the recent low value of the solar oxygen abundance still 
exceeds slightly the present-day total oxygen abundance in the interstellar 
medium at the solar galactocentric distance (see also Daflon et al. 2001). 
If it is the case, this is evidence that the variation of oxygen abundance at 
the solar galactocentric distance with time is not smooth (for example, due to 
an irregular rate of unenriched gas infall onto the disk, see Pilyugin 
\& Edmunds 1996). The discussion of the model for the 
chemical evolution of our Galaxy is far beyond the goal of the present study,
we only would like to note that the recent low solar oxygen abundance results in 
better agreement with the present-day total oxygen abundance in the interstellar 
medium (and in young stars) at the solar galactocentric distance than 
generally accepted earlier solar values. 

\section{Shaver et al.'s set of H\,{\sc ii} regions}

Combining radio and optical spectroscopy Shaver et al. (1983) have found 
chemical abundances in a large sample of Galactic H\,{\sc ii} regions covering 
a wide range in galactocentric radius. Radio recombination lines have been 
used to determine electron temperatures in H\,{\sc ii} regions and these 
temperatures have been applied to optical spectra of the same H\,{\sc ii} 
regions to determine the abundances of oxygen and other elements. 
Shaver et al. (1983) have found that their oxygen abundances are in agreement 
with the empirical R$_{23}$ -- calibration. At the same time it has been found 
(Pilyugin 2000, 2001a,b, 2003) that the oxygen abundances derived with the R$_{23}$ -- 
calibration after Edmunds \& Pagel (1984) involve a systematic error depending 
on the excitation parameter: the R$_{23}$ -- method provides more or less 
realistic oxygen abundances in high-excitation H\,{\sc ii} regions and yields an 
overestimated oxygen abundances in low-excitation H\,{\sc ii} regions. Taking 
into account that many H\,{\sc ii} regions in the sample of Shaver et al. (1983) 
are low-excitation objects, the agreement of Shaver et al's oxygen abundances 
with the empirical R$_{23}$ -- calibration hints that their oxygen abundances 
are overestimated. 

\begin{table}
\caption[]{\label{table:shav}
Oxygen abundances in set of H\,{\sc ii} regions from Shaver et al. (1983). 
The name of H\,{\sc ii} region is given in column 1. The galactocentric distance 
taken from Deharveng et al. (2000) is reported in column 2.
The original oxygen abundance (O/H)$_{SH}$ from Shaver et al. (1983) is 
listed in column 3.  
The oxygen abundance (O/H)$_{P}$ determined here with the P -- method 
is reported in column 4. 
}
\begin{center}
\begin{tabular}{lccc} \hline \hline
             &              &                   &                  \\
             &galactocentric& (O/H)$_{SH}^*$    &  (O/H)$_P^*$     \\
H\,{\sc ii} region   & distance     & Shaver et al.     &  this work  \\  
             & (kpc)        & (1983)            &                  \\  \hline
NGC6604-1    &     6.5      &    9.11           &    8.61          \\
NGC6604-2    &     6.5      &    8.72           &    8.56          \\
M16          &     6.6      &    8.76           &    8.63          \\
l Cen-2      &     7.8      &    8.79           &    8.66          \\
RCW40        &     8.7      &    8.85           &    8.44          \\
RCW34        &     9.2      &    8.88           &    8.56          \\
RCW19        &     9.6      &    8.64           &    8.24          \\
Rosette-1    &    10.0      &    8.47           &    8.32          \\
S252-1       &    10.7      &    8.34           &    8.62          \\
RCW16-1      &    10.9      &    8.69           &    8.42          \\
RCW8         &    11.3      &    8.65           &    8.27          \\
RCW6         &    11.6      &    8.69           &    8.51          \\
RCW5-1       &    13.0      &    8.33           &    8.34          \\
G201.6+1.6   &    14.0      &    8.55           &    8.04          \\
S284         &    14.6      &    8.36           &    8.27          \\ \hline 
\end{tabular}
\end{center}
$^*$  in units of 12+log(O/H)
\end{table}

\begin{figure}
\centering
\resizebox{\hsize}{!}{\includegraphics[angle=0]{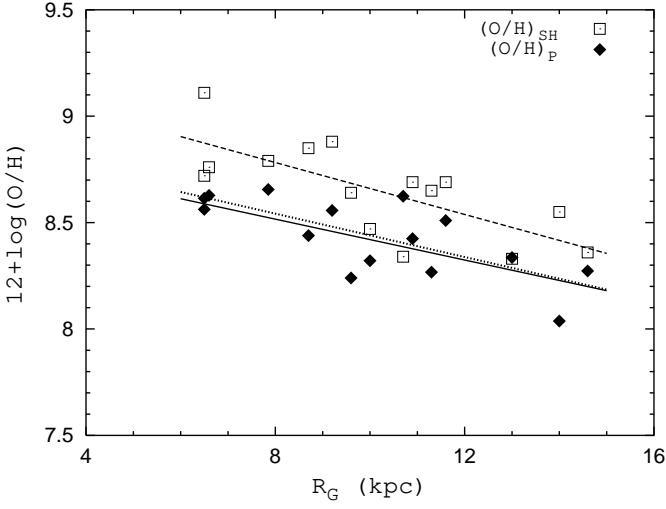}}
\caption{
Oxygen abundance in Galactic H\,{\sc ii} regions from sample of Shaver et al. 
(1983) as a function of galactocentric distance. 
The galactocentric distances are from Deharveng et al. (2000). 
The open squares are original oxygen 
abundances (O/H)$_{SH}$ from Shaver et al. (1983), the dashed line is the 
best fit to those data. The filled rhombuses are oxygen 
abundances (O/H)$_P$ derived here with the P -- method using the same 
measurements of the line intensities, the dotted line is the best fit to 
those data. The solis line is the (O/H)$_{Te}$ -- R$_G$ relationship (see 
Fig.\ref{figure:ohte}).
}
\label{figure:ohp}
\end{figure}

Recently a new way of oxygen abundance determination in H\,{\sc ii} regions, 
in which the physical conditions in H\,{\sc ii} region are estimated via the excitation 
parameter P (P -- method), was suggested (Pilyugin 2001a). 
By comparing the oxygen abundances in high-metallicity H\,{\sc ii} regions derived 
with the $T_{e}$ -- method and those derived with the P -- method, it was found 
that the precision of oxygen abundance determination with the P -- method is 
comparable to that of the $T_{e}$ -- method (Pilyugin 2001a,b). 
Then the comparison of oxygen abundances obtained by Shaver et al. (1983) and 
oxygen abundances determined with the P -- method using the same line intensities 
can give some idea of the accuracy of Shaver et al's oxygen abundances. 
The oxygen abundances obtained by Shaver et al. (1983) will be referred to as 
(O/H)$_{SH}$ abundances, the oxygen abundances derived with the P  -- method 
will be referred to as (O/H)$_{P}$ abundances. Only H\,{\sc ii} regions with 
reliable measured ratios [OII]$_{3727,3729}$/H$_{\beta}$ and 
[OIII]$_{4959,5007}$/H$_{\beta}$ from the sample of Shaver et al. (1983) will be 
considered. These H\,{\sc ii} regions are listed in Table \ref{table:shav}.
(O/H)$_{SH}$ abundances as a function of galactocentric distance are shown 
in Fig.\ref{figure:ohp} by the open squares. The linear best fit to those data
is presented by the dashed line in Fig.\ref{figure:ohp}. The galactocentric 
distances are from Deharveng et al. (2000). The best fit to (O/H)$_{SH}$ 
abundances is given by the following equation 
\begin{equation}
12 + \log (O/H)_{SH} = 9.27 - 0.061 \, R_G  .
\label{equation:gradshav}
\end{equation}
The scatter is equal to 0.15 dex. Eq.(\ref{equation:gradshav}) results in the 
oxygen abundance at the solar galactocentric distance as large 
as 12+log(O/H) = 8.75. 

The (O/H)$_{P}$ oxygen abundances in H\,{\sc ii} regions are determined with 
the expression suggested in (Pilyugin 2001a)
\begin{equation}
12+log(O/H)_{P} = \frac{R_{23} + 54.2  + 59.45 P + 7.31 P^{2}}
                       {6.07  + 6.71 P + 0.37 P^{2} + 0.243 R_{23}}  ,
\label{equation:ohp}
\end{equation}
where $R_{23}$ =$R_{2}$ + $R_{3}$, 
$R_{2}$ = $I_{[OII] \lambda 3727+ \lambda 3729} /I_{H\beta }$, 
$R_{3}$ = $I_{[OIII] \lambda 4959+ \lambda 5007} /I_{H\beta }$, 
and P = $R_{3}$/$R_{23}$. 
The (O/H)$_{P}$ abundances in H\,{\sc ii} regions from the sample of Shaver 
et al. (1983) are listed in column 4 in Table \ref{table:shav}.
(O/H)$_{P}$ abundances as a function of galactocentric distance are shown 
in Fig.\ref{figure:ohp} by the filled rhombuses. The linear best fit to those 
data is presented by the dotted line in Fig.\ref{figure:ohp}.
The best fit to (O/H)$_{P}$ abundances is given by the following equation 
\begin{equation}
12 + \log (O/H)_P = 8.95 - 0.051 \, R_G      .
\label{equation:gradohp}
\end{equation}
The scatter is equal to 0.12 dex. Eq.(\ref{equation:gradohp}) results in the 
oxygen abundance at the solar galactocentric distance as large 
as 12+log(O/H) = 8.52. 

Examination of Fig.\ref{figure:ohp} (see also the data from Table \ref{table:shav}) 
shows that there is a systematic difference between (O/H)$_{SH}$ and (O/H)$_{P}$ 
abundances, (O/H)$_{SH}$ abundances are higher on average by 0.2-0.3 dex than 
(O/H)$_{P}$ abundances. Taking into account that the precision of oxygen 
abundance determination with the P -- method is comparable to that of the 
$T_{e}$ -- method (Pilyugin 2001a,b) we can conclude that  the Shaver et al.'s 
oxygen abundances are overestimated by 0.2-0.3dex.

Our conclusion can be verified by comparison of the radial (O/H)$_{SH}$ 
abundances distribution with the radial (O/H)$_{T_{e}}$ abundance distribution
across the disk of the Milky Way Galaxy. 
(It should be stressed that the reliability of (O/H)$_{T_e}$ abundances in 
H\,{\sc ii} regions is confirmed by the agreement between the value of the 
oxygen abundance at the solar galactocentric distance traced by (O/H)$_{T_e}$ 
abundances in the H\,{\sc ii} regions and that derived from the interstellar 
absorption towards stars (a model-independent method).) The (O/H)$_{Te}$ -- 
R$_G$ relationship obtained above is shown in Fig.\ref{figure:ohp} by a solid 
line. Inspection of Fig.\ref{figure:ohp} shows that the (O/H)$_{SH}$ abundances 
lie above (in average by 0.2-0.3 dex) the (O/H)$_{Te}$ -- R$_G$ relationship. 
This confirms our conclusion that the Shaver et al.'s oxygen abundances are 
overestimated by a factor of about 2. 

\section{Conclusions}

The compilation of spectra of Galactic H\,{\sc ii} regions with available 
diagnostic [OIII]$\lambda$4363 line was carried out. Our list contains 71 
individual measurements of 13 H\,{\sc ii} regions in the range of 
galactocentric distances from 6.6 to 14.8 kpc. The oxygen abundances in 
all the H\,{\sc ii} regions were recomputed in the same way, using the classic 
T$_e$ - method. 

The oxygen abundance at the solar galactocentric distance traced by H\,{\sc ii} 
regions is in agreement with the oxygen abundance in the interstellar medium 
in the solar vicinity derived with high precision from the interstellar 
absorption lines towards stars. 

The obtained value of the central oxygen abundance in the disk 
of the Milky Way Galaxy 12 + log(O/H)$_{Te}$(R=0) = 8.90 lies in the same 
range as the values of the central oxygen abundance in the disks 
of other spiral galaxies, in which the oxygen abundances in H\,{\sc ii} 
regions were derived with the $T_{e}$ -- or with the P -- method. 

The derived radial oxygen abundance distribution was compared with that for 
H\,{\sc ii} regions from the Shaver et al. (1983) sample which is the basis of 
many models for the chemical evolution of our Galaxy. It was found that the 
original Shaver et al.'s oxygen abundances are overestimated by 0.2-0.3 dex. 
Oxygen abundances in H\,{\sc ii} regions from the Shaver et al. sample have been 
redetermined with the recently suggested P -- method. The radial distribution of 
oxygen abundances from the Shaver et al. sample redetermined with the P -- method 
is in agreement with the radial distribution of O/H$_{T_{e}}$ abundances 
obtained here.

\begin{acknowledgements}
We thank the anonymous referee for helpful comments. This study was partly 
supported by the Joint Research Project between Eastern Europe and Switzerland 
(SCOPE) No. 7UKPJ62178 (L.S.P.), the NATO grant PST.CLG.976036 (L.S.P. and F.F.), 
the Italian national grant delivered by the MURST (L.S.P.). 
\end{acknowledgements}

\end{document}